\documentclass[prl,nofootinbib,superscriptaddress,twocolumn]{revtex4}

\def\mysections#1{{\bf #1.} }

\usepackage{amssymb}
\usepackage{amsmath}
\usepackage{graphicx}
\usepackage{longtable}
\usepackage{verbatim}
\usepackage{amsfonts}

\newcommand{\be}{\begin{equation}}
\newcommand{\ee}{\end{equation}}
\newcommand{\bea}{\begin{eqnarray}}
\newcommand{\eea}{\end{eqnarray}}
\newcommand{\beq}{\begin{equation}}
\newcommand{\eeq}{\end{equation}}
\def\beqa{\begin{eqnarray}}

\def\eeqa{\end{eqnarray}}

\def\lsim{\mathrel{\rlap{\lower4pt\hbox{\hskip0.5pt$\sim$}}
    \raise1pt\hbox{$<$}}}         
\def\gsim{\mathrel{\rlap{\lower4pt\hbox{\hskip0.5pt$\sim$}}
    \raise1pt\hbox{$>$}}}         

\begin{document}

\vspace*{-30mm}

\title{ A Possible Cosmological Explanation of why\\
\vspace*{0.2cm}
Supersymmetry is hiding at the LHC}

\vspace*{2cm}
\author{ Antonio Riotto}
\affiliation{D\'epartement de Physique Th\'eorique and Centre for
  Astroparticle Physics (CAP),\\
  \vspace*{0.1cm}
Universit\'e de Gen\`eve, 24 quai E. Ansermet, CH-1211 Gen\`eve, Suisse}
\vspace*{1cm}

\begin{abstract}
\noindent
If one is not ready to pay a large fine-tuning price within supersymmetric models given the current measurement of the Higgs boson mass, one can envisage a scenario where the supersymmetric spectrum is made of heavy scalar sparticles and much lighter fermionic superpartners.
We offer a  cosmological explanation of why nature might have chosen such a mass pattern:  the opposite mass pattern
 is not observed experimentally because it is not compatible 
  with the 
   plausible idea that the universe went through a period of primordial inflation.

%
%
\end{abstract}

\maketitle

\noindent
\noindent
There are many good reasons to believe that the Standard Model (SM)  is not the ultimate
theory  of nature since it is unable to answer many fundamental questions. One of them,
why and how the electroweak scale and the Planck scale are so hierarchically separated
has motivated the Minimal Supersymmetric extension of the Standard Model (MSSM) as
the underlying theory at scales of order the TeV. 
However, the CERN collaborations  ATLAS \cite{ATLAS} and CMS \cite{CMS} have recently reported the discovery of a boson,  with  mass 
around $(125-126)$  GeV, whose branching ratios are  fully consistent with   
the SM Higgs boson. 

Such a value of the Higgs mass implies, in the context of the MSSM,   a top squark heavier than a  few TeV, which in turn  causes a
fine-tuning at the   percent level. Therefore, 
the lack of evidence for sparticles at LHC  suggests that either low-energy supersymmetric  theories are
fine-tuned or (some of the) sparticles are much heavier than the weak scale. This would leave gauge coupling unification as the only, albeit indirect,  evidence
for low-scale supersymmetry \cite{un} (besides the presence of dark matter provided by the lightest neutralino). 

On the other hand,  gauge coupling unification can be achieved  for heavy scalar sparticles and much lighter fermionic superpartners  (gauginos and
higgsinos). For instance, choosing the fermion masses near a TeV, as dictated by reproducing the correct dark matter abundance,
reproduces successful unification independently of the masses of scalar sparticles. This is the underlying idea 
of split supersymmetry \cite{split} where  sfermion masses can be as large as the unification scale and, more recently of mini-split supersymmetry \cite{minisplit}, where the mass of the Higgs correlates with the mass of the sfermions which have to be in the $(10-10^5)$  TeV range \cite{corr} (unless $\tan\beta\lsim 3)$.

The question we would like to ask in this short note is: why is supersymmetry hiding in  LHC searches or, put in more modest way, why are sfermions masses $m$  so much heavier than the gaugino and higgsino masses $M\sim {\cal O}(1-10)$ TeV? The  answer we wish to offer is based on cosmological considerations: the opposite mass pattern
 is not observed experimentally because it is not compatible with the universe as we know it.
 The logic  relies on  two well-accepted ideas.

 First, there are many (over a hundred) flat directions in the
field space of the MSSM. It may happen that some combination of the squark and/or
slepton mass-squared parameters get negative at some scale  below the Planck scale
when running through the Renormalization Group Equations (RGE's) from the weak
scale up. This may happen if the sfermion masses are lighter than the gaugino masses,  leading either to the appearance of unacceptable color/charge breaking minima (provided that  nonrenormalizable superpotentials are present which can lift the direction)
or to unbounded from below directions in the effective potential for the squark and/or
slepton fields.  

The situation here is fairly analogous to what happens for the effective potential of the
SM Higgs field: for a top quark mass large compared to the Higgs and gauge bosons
masses, the one-loop top quark contribution to the effective potential will dominate the
others and drive the coefficient of the quartic term  negative for very large values of the Higgs field
thus destabilizing the effective potential and making our vacuum
a local, but not global, minimum \cite{higgs1,higgs2,higgs3,higgs4}.

Secondly, one of the basic ideas of modern cosmology is that there was an epoch early
in the history of the universe when potential, or vacuum, energy 
dominated other forms of energy density such as matter or radiation. 
During such a
vacuum-dominated de Sitter stage,  the scale factor grew (nearly) exponentially
in time. During this phase, dubbed inflation 
\cite{guth81,lrreview},
a small,  smooth spatial region of size less than the Hubble radius
could grow so large as to easily encompass the comoving volume of the 
entire presently observable universe. If the universe underwent
such a period of rapid expansion, one can understand why the observed
universe is so homogeneous and isotropic to high accuracy.
Inflation has also become the dominant 
paradigm for understanding the 
initial conditions for the large scale structure  formation and for cosmic
microwave background  anisotropy.  

A key feature of inflation is that, during a period of de Sitter characterized by a nearly constant Hubble rate $H$, 
all scalar fields with 
 mass smaller than the Hubble rate $H$  are inevitably quantum-mechanically excited
with a nearly flat spectrum. The consequence of it is that  excessive fluctuations of the scalar field parametrizing the flat direction during
inflation would  destabilize
our current  vacuum. If the Hubble rate is large enough during inflation and the
sfermion masses are too light, then the classical value of the flat directions field  is pushed
above their instability point towards  color/charge breaking minima which would either not correspond to the observable universe
we live in (and in fact this reasoning would extend to the whole inflated region, much beyond our observed universe) or stop inflation after a few e-folds. Both possibilities are not acceptable.
One can therefore reasonably conclude that the reason why   sfermion masses are heavier than
gaugino masses is because the opposite situation is not compatible with the idea that the universe suffered a period of inflation without
destabilizing our well-behaving vacuum.

Let us elaborate now a bit further on these ideas and let us consider, out of the many MSSM flat directions,   the $u_1d_2d_3$ (subindices indicate the quark family) flat direction parametrize by the field $\phi$. Along this particular direction (as for many others) the coefficient 
 of the quartic term $\phi^4$ is vanishing for all scales 
$\mu$, and the renormalization group improved potential reads
%

\begin{equation}
\label{eff}
V(\phi)=\frac{1}{2}m^2(\phi)\phi^2,
\end{equation}
where we have conveniently set the renormalization scale $\mu=\phi$ in order to  make the one-loop logarithms
small. 
The RGE for $m^2$ is given by 
\begin{eqnarray}
\mu\frac{{\rm d} m^2}{{\rm d} \mu}&=&\frac{1}{8\pi^2}\left[-16g_3^2 M_3^2-
\frac{8}{3}\:g_1^2M_1^2\right.\nonumber\\
&+&\left. 2h_b^2\left(m^2_{\widetilde{q}_L}+m^2_{\widetilde{b}_R}+
m^2_{H_1}+A_b^2\right)\right],
\label{full}
\end{eqnarray}
where $g_1$ is the standard $U(1)$ 
coupling, $M_i$ are gaugino masses, $h_b$ is the 
bottom quark Yukawa coupling and $A_b$ is the bottom quark 
trilinear mixing parameter, and all parameters are running. 
If $\tan\beta$ (the ratio of the two neutral MSSM
Higgs vacuum expectation values) is not too large,
$h_b$ is small and the term proportional to $h_b^2$ in the above 
equation  can be neglected. Furthermore, if one takes  $M_1\leq M_3$, as usually results
in GUT models, the effects of the term proportional to $g_1^2$ in
Eq. (\ref{mq.eq}) is also negligible. This leads  to the solution  
\begin{eqnarray}
m^2(\mu)&=&m^2-\frac{2}{\pi^2}g_3^2(M_3)M_3^2{\rm ln}(\mu/M_3)\nonumber\\
&\times&
\left\{\frac{1+3 g_3^2(M_3){\rm ln}(\mu/M_3)/(16\pi^2)}{
\left[1+3 g_3^2(M_3){\rm ln}(\mu/M_3)/(8\pi^2)\right]^2}\right\},
\nonumber\\
&&
\label{mq.eq}
\end{eqnarray}
where all masses on the right-hand side are physical 
(propagator pole) masses evaluated at the low-energy scale $M_3$ and the only undertermined factors 
is $g_3^2(M_3)$, which depends on the full spectrum 
of the superpartner masses and the initial condition 
$\alpha_3(M_Z)=0.12$. To obtain $g_3(M_3)$ we have assumed a
common supersymmetric threshold at $M_3$, but the results are 
almost unsensitive to this simplifying assumption.
Eq. (\ref{mq.eq}) delivers an instability   scale 
\begin{equation}
\Lambda\simeq M_3\:{\rm exp}\left[ {\pi^2m^2\over 2g_3^2M_3^2}\right],
\label{phi0.eq}
\end{equation}
technically defined to be the scale at which the potential has a maximum.
Solving Eq. (\ref{full}) numerically without approximations, it turns out that  for $m\lsim  0.7 M_3$ \cite{falk,rr} the  quadratic mass of the flat direction can run to negative values, signalling
the appearance of an instability at some energy value $\Lambda$ smaller than the GUT scale. 
We find that $\Lambda\simeq 10^{10}\,(10^{6})$ GeV for $m\simeq 600$ GeV and $M_{3}\simeq$ 1 (2) TeV. It is also clear from 
Eq. (\ref{mq.eq}) that in the opposite case, $m\gsim M_3$, no instability occurs. 

The color conserving minimum,
although metastable, has a lifetime longer than the present age of
the universe and can survive both quantum tunneling and the effects of
high temperatures in the early universe, causing the color/charge breaking
effects to be in practice not dangerous \cite{rr}. This holds in the post-inflationary stage though. 

What may happen during inflation
if  sfermion masses $m$ are lighter than gaugino masses $M$?  As we mentioned above,  a period of inflation  destabilizes the (color/charge preserving) vacuum and may even stop inflation.
The process of generating a classical flat direction  field $\phi$ configuration
in the inflationary universe can be interpreted as the
result of the Brownian motion of the  field $\phi$ under the
action of its quantum fluctuations which are converted into
the classical field when their wavelengths overcome the Hubble
length. 

The best way to describe the structure of the  fluctuations 
is provided by the stochastic approach in which one defines
the comoving distribution of probability $P_c(\phi,t)$ to find the  field value 
 at a given time at a given point \cite{star}. 
The subscript $c$ serves to indicate that $P_c$ corresponds to the 
fraction of original comoving volume filled by the flat direction  field
 at the time $t$. The comoving probability 
satisfies the Fokker-Planck equation
\begin{equation}
\label{fp}
\frac{\partial P_c}{\partial t}=\frac{\partial}{\partial \phi}\left[
\frac{H^3}{8\pi^2}\frac{\partial P_c}{\partial \phi} 
+\frac{V'(\phi)}{3H}P_c\right].
\end{equation}
To solve this equation  we assume that 
the Hubble rate is approximately constant during inflation. Furthermore, 
we assume that the flat direction field
is initially localized at $\phi=0$, $P_c(\phi,0)=\delta_D(\phi)$, and study its evolution. 
Eq. (\ref{fp}) can be solved by separation of variables
\beq
P_c(\phi,t)=\sum_{n=0}^\infty c_n e^{-\left( \alpha 
V+\frac{H^3a_n t}{8\pi^2}\right)}g_n(\phi) ,
~~~~~\alpha=\frac{8\pi^2 }{3H^4}\ ,
\label{solexp}
\eeq
where $g_n$ and $a_n$ are the eigenfunctions and the eigenvalues of 
the equation $
g_n^{\prime\prime}-\alpha V^\prime g_n^{\prime}=-a_n g_n$.
We are interested in the case in which the Hubble rate 
is much larger than $\Lambda$. In this case the potential term  can be neglected since $\alpha V^\prime 
g^\prime_n/g_n^{\prime\prime} \lsim (m^2\Lambda^2 /H^4) \ll 1$, 
once we use the condition $\phi<\Lambda$.
One can  easily get convinced  that, for $H\gg \Lambda$,  
the variance of the field is given by $\langle\phi^2\rangle\sim (H^3t/4\pi^2)$. This means that after a short time $t\sim \Lambda^2/H^3\ll H^{-1}$, the typical value of the field is of the order $\Lambda$ and the instability region is easily accessible (this happens much before  the evolution
of the flat direction might be blocked by other non-flat direction \cite{dani}). This is also reflected in the computation of the survival probability $P_\Lambda$ for the flat direction field to remain in the 
region $\phi<\Lambda$: it  is  exponentially suppressed at times $t\gg H^{-1}$ (for technicalities see Ref. \cite{higgs1})
\beq
P_\Lambda(t)\equiv \int_{0}^{\Lambda}\, {\rm d}\phi~ P_c(\phi,t)  \simeq \frac{2}{\pi} e^{-\frac{H^3t}{32\Lambda^2}},\,\,\,(H\gg \Lambda).
\eeq
This is valid for $t \gg \Lambda^2/H^3$ 
and therefore it happens very rapidly, justifying the assumption of 
neglecting the time dependence of the Hubble constant.

Since most models of inflation predict a  large number of e-folds $N\simeq Ht$, we conclude that a de Sitter stage with 
Hubble rate $H$ larger than $\Lambda$ will lead to a current universe which looks completely different from ours if sfermion masses are lighter than gaugino masses:  a universe where
either color and charge are broken or a universe where inflation took place only for a few e-folds before the negative  energy density of the runaway direction  took over the vacuum energy density driving inflation. It is indeed important to keep in mind that truly runaway directions may exist: because of the non-renormalization theorem for the superpotential,
even operators consistent with all symmetries need not appear in the superpotential; 
in certain instances the absence of the gauge invariant operators which could lift the flat
direction can be guaranteed by an $R$ symmetry;  directions of this type are therefore exactly
flat in the supersymmetric limit. Only soft terms contribute to the potential along
such  runaway directions.

We acknowledge that -- so far --  our logic has  two weak points: 

\noindent
1) we have assumed  that the Hubble rate during inflation is larger than the would-be instability scale, $H\gg \Lambda$. 
It is interesting to note that the Hubble rate 
parametrizes 
the amount of tensor perturbations during inflation. Tensor modes  can give rise to $B$-modes
of polarization of the CMB radiation
through Thomson scatterings of the CMB photons off free electrons
at last scattering \cite{pol}. The amplitude of the $B$-modes
depends on the amplitude of the gravity waves
generated during inflation, which in turn depends on the
energy scale at which inflation occured. Current CMB anisotropy data impose the upper bound on the tensor-to-scalar power ratio
$T/S\lsim 0.5$ \cite{wmapping} and 
PLANCK's expected sensitivity is  about $T/S=0.05$. This  corresponds to a minimum testable value of    $H\simeq 
6.7\times 10^{13}$ GeV \cite{plancktensor}. A detection of tensor modes by   PLANCK would indicate a large value of the Hubble rate during 
inflation and would therefore support  the logic put forward in this note; 

\noindent
2) it explains why  the  hierarchy $m\lsim  M$ may  not be  cosmologically acceptable,  but not why $m\gg M$. 

In particular, having sfermions as massive as  fermionic superpartners, $m\sim M$, seems legitimate from the cosmologically point of view.
How can we explain the large hierarchy $m\gg M$? So far, we have assumed the validity of the form (\ref{eff}) for the flat direction potential. During inflation the potential of the flat direction might not be of the simple form (\ref{eff}) and  mass squared of the flat directions 
of the form ${\cal O}(H^2)$ may be 
generated from the corrections to the K\"{a}hler potential 

\be
\delta {\cal K}\supset\int\,{\rm d}\theta\, \frac{\chi^\dagger\chi}{M_{\rm pl}}\, \phi^\dagger\phi\supset c_m H^2\phi^2,
\ee
where $\chi$ is a field which dominates the energy density of the universe,  $M_{\rm pl}$ is the reduced Planck mass and $c_m$ is a ${\cal O}(1)$ coefficient  \cite{susyinf}. 
 Suppose
the parameter  $c_m$ is negative for at least one dangerous flat direction. This is not  inconceivable as 
a positive contribution in the   K\"{a}hler  potential gives a negative contribution to $m^2$. If so,    our logic is  strengthened: 
 if   at least one of the many flat directions whose mass squared runs to negative values at some scale  below the Planck scale has $c_m<0$ and is   a truly runaway direction, then the only safe possibility  is that  the mass of the sfermion at the low-energy scale is large

\be
\label{mH}
m^2\gsim H^2\gg M^2 \sim{\rm TeV}^2,
\ee
otherwise   inflation is promptly stopped and $M\sim$ TeV is the condition to have a successful
dark matter candidate. 
This conclusion is further fortified by the fact that also gaugino masses can obtain ${\cal O}(H)$ corrections to their masses, thus
leading to a faster  running of $m^2$ towards negative values. There is no indication that the  Hubble parameter may not be
in the $(10-10^5)$  TeV range. If so,  the condition (\ref{mH}) might be a natural explanation for the mini-split spectrum preferred by the recent data on the Higgs mass. If, on the other hand, tensor modes will be soon discovered signaling a high energy scale for inflation, then the split supersymmetric spectrum will be
favoured. 

The condition (\ref{mH}) may be   necessary even when $m\gg M$ if one considers the fact that  multipole flat directions are present. Suppose that 
$M\ll m\ll H$ and $c_m<0$ for  a flat direction which this time is lifted by a nonrenormalizable superpotential
 containing a single field $\psi$ not in the flat direction and some number of fields which
make up the flat direction 

\be 
{\cal W}_n=\frac{\lambda}{M^{n-3}}\psi\phi^{n-1},
\ee
for some mass scale $M$. Examples of this type are represented by the flat direction $udd$ which might be  lifted by ${\cal W}_4=uude/M$ and by the $Que$ direction possibly lifted by  ${\cal W}_9=QuQuQuH_1ee/M^6$. When the flat direction gets a vacuum expectation value $\langle\phi\rangle\sim\left(c_m H^2 M^{2(n-3)}/\lambda^2\right)^{1/(2n-4)}$, then the flat direction involving the $\psi$ field ({\it e.g.} the $LLe$ direction
for ${\cal W}_4$) will be destabilized by the associated tadpole, even if their corresponding mass squared ${\cal O}(H^2)$ is positive

\be
V(\psi)=+H^2\psi^2-\lambda H\psi\langle\phi\rangle^{n-1}/M^{n-3}+\cdots,
\ee
where the dots indicate higher order terms. The tadpole leads to a  negative energy density $\sim
M^4\left(H^2/M^2\right)^{(n-1)/(n-2)}/\lambda^{2/(n-1)}$. This energy can be easily larger than the de Sitter  energy density $\sim H^2M_{\rm pl}^2$, thus stopping inflation, if for instance $M=M_{\rm pl}$ and $\lambda\lsim(H/M_{\rm pl})^{(n-1)/(n-2)} $. The condition (\ref{mH}) should be invoked to avoid such a  disaster. 

If $0<c_m\ll 1$, the same conclusion can be reached. In such a case the de Sitter induced fluctuations of the flat direction can be as large as 
$\langle\phi^2\rangle\sim 3H^4/8\pi^2 m^2\sim 10^{-1} H^2/c_m^2$. This fluctuation induces a tadpole in the potential of the other
flat direction $\psi$ and the  resulting negative energy density is larger than the vacuum energy density driving inflation if
$c_m\lsim (H/M_{\rm pl})^{(n-2)/2(n-1)} $. For $n$ large, this is not a tight constraint and once again  the condition (\ref{mH}) 
can save the situation.

In most of the literature flat direction vacua during inflation are assumed  to have  an energy density smaller  than the one driving inflation. 
In this short note we have argued that --  for at least one flat direction --  this assumption is not correct, then the hierarchy $m\gg M$ might be 
 motivated from the cosmological point of view: once inflation is accepted, large sfermion masses ensure the stability of the vacuum and the universe with all the LSS as we know it. In the opposite case $m\lsim M$ either inflationary quantum mechanical fluctuations or 
 ${\cal O}(H^2)$ corrections to the sfermion masses may be harmful.  
  It would be an indication that nature has been kind to us: she
has provided a primordial period of inflation to generate the structures we observe and live in,  she has  hidden sufficiently well supersymmetry, still providing us   with a livable vacuum,    and she has provided light enough supersymmetric fermions to explain the dark matter puzzle. 

%
%

\vskip 0.5cm
\noindent
\mysections{Acknowledgments}
We thank S. Dimopoulos, A. De Simone, J.R. Espinosa, D. Figueroa,  G.F. Giudice,  A. Kehagias, M. Maggiore, L. Senatore  and M. Sloth for reading the manuscript and for providing useful comments, criticisms and suggestions.  
The author  is supported by the Swiss National
Science Foundation (project number  200021140236). 


\end{document}